# Low-temperature Raman spectra of L-histidine crystal

G.P. De Sousa, P.T.C. Freire[*], J. Mendes Filho, F.E.A. Melo

Departamento de Física, Universidade Federal do Ceará
Campus do Pici, C.P. 6030 Fortaleza – CE 60455-760 Brazil

C.L. Lima

Departamento de Física, Campus Ministro Petrônio Portella,
Universidade Federal do Piauí, 64049-550, Teresina-PI, Brazil

Abstract

We present a Raman spectroscopy investigation of the vibrational properties of L-histidine crystals at low temperatures. The temperature dependence of the spectra show discontinuities at 165 K, which we identify with modifications in the bonds associated to both the $NH_3^+$ and $CO_2^-$ motifs indicative of a conformational phase transition that changes the intermolecular bonds. Additional evidence of such a phase transition was provided by differential scanning calorimetry measurements, which identified an enthalpic anomaly at ~165 K.

Introduction:

Amino acids are organic molecules characterized by carboxylic and amino groups with the general formula $NH_2$-CH(R)-COOH, where R is a radical characteristic of the molecule. In aqueous solution, for most pH values, and in the solid state, the amino acids take a zwitterionic form, i.e., $NH_3^+$-CH(R)-$COO^-$. Three special features of the amino acids have attracted attention to the chemical and physical properties of their crystals: they (i) serve as biomimetics, modeling interactions in biopolymers [1], (ii) present a rich phase diagram, a facet closely related to polymorphism, with applications in pharmaceutical industry [2], and (iii) contain an infinite sequence of head-to-tail hydrogen bonds, which gives controlled access to a chemical bond that plays leading roles in catalytic biochemical processes [3]. Such studies are expected to yield important background information concerning both the static structure of proteins and such dynamical properties as denaturation, renaturation, folding, and changes in folds.

Among the amino acids, L-histidine has raised special interest both because it has a variety of functionalities found in biological systems and because the zwitterion displays two tautomeric forms occurring in equilibrium at biological pH [4]. The L-histidine crystal is moreover characterized by an imidazole ring, formed by two nitrogen and three carbon atoms, as depicted in the inset of Figure 1. In solution, histidine presents five pH-dependent different protonation states [5,6] and shows six distinct geometrical conformations [5]. As a consequence, histidine can act as a donor or as an acceptor for the hydrogen bond [7], influencing the secondary and tertiary structure of proteins and participating in several enzyme reactions [8]. In particular, the imidazole of histidine participates in the tuning of the electronic properties of the copper ion, which is involved in catalysis [8, 9]. Given these features, many studies focusing on different aspects of bond formation and intermediation by histidine have been reported [10–12].

Temperature-dependent studies of the physical properties, such as piezoelectrical or nonlinear optical properties, of amino-acid crystals can provide information on their contribution to the dynamic properties of the proteins [13, 14]. Decreasing temperatures and increasing pressures tend to reduce the crystal-lattice parameters. In addition, temperature changes also modify the phonon population. Temperature decrements and pressure increments may differently reorient the molecules relative to each other and to the structure of hydrogen-bonded networks [15]. In brief, to understand all possible conformations of molecules, different hydrogen bond networks and, in certain cases, new polymorphs of a certain structure, one must study the material under different extreme pressure or temperature conditions.

Previous low-temperature studies of amino-acid crystals have yielded interesting results. For example, both the intramolecular conformations and the intermolecular hydrogen bonding in L-cysteine were found to change [16]. While at room temperature SH…O hydrogen bonds dominates over SH…S bonds, the inverse occurs at low temperatures. The transition from one phase to the other undergoes different conformation changes, several molecular fragments being activated at different temperatures. For the L-histidine hydrochloride monohydrate crystal [17], the emergence of additional modes in the low-wavenumber region of the Raman spectrum upon cooling has revealed a structural phase transition between 140 and 110 K.

Here we ask whether L-histidine shows transitions analogous to those observed in the L-cysteine or L-histidine hydrochloride monohydrate crystal. To answer this question, we have measured Raman spectra of the L-histidine crystal under low-temperature conditions, between 30 and 300 K. While most modes are weakly temperature dependent, a few show anomalies pointing to a conformational change at ~160 K. As a check, we have carried out differential scanning calorimetry (DSC) measurements, the results of which are consistent with the Raman spectra. We analyze the results of the two experimental techniques to provide a simple interpretation of the observed instability.

Experimental:

Our study used polycrystalline samples of commercially-available L-histidine, from Vetec Quimica Fina. Powder X-ray diffraction measurements with a DMAXB Rigaku diffractometer using Cu $K_\alpha$ radiation operating at 40 kV, 25 mA confirmed the crystal structure. The Raman spectra of L-histidine were obtained with a T64000 Jobin Yvon spectrometer and an argon ion laser operating at 514.5 nm as exciting source. The slits of the spectrometer allowed a resolution of ~ 2 cm$^{-1}$. Low-temperature measurements were carried out in a helium-flux cryostat, a digital temperature controller with 0.1K stability monitoring the temperature in the 30 – 285 K interval. The spectra were measured following an appropriate thermal stabilization time after each thermal step. Differential scanning calorimetry (DSC) measurements were made with a Netzsch Instrument (DSC 204 F1 – Phoenix) equipped with a liquid nitrogen cooling accessory. A ~3mg sample was monitored between 300K and 140 K, with a 5 K/min cooling rate.

Results and Discussion:

L-histidine can crystallize in two polymorphic forms, monoclinic [18] or orthorhombic [19]. At ambient conditions, X-ray diffraction measurements (Fig. 1) showed that the L-histidine in our experiments had crystallized in the orthorhombic $P2_12_12_1$ ($D_2^4$) structure, with four molecules of $C_6H_9N_3O_2$ (LHIS) per unit cell and lattice parameters a = 5.177, b = 7.322 and c = 18.87 Å [19]. The L-histidine molecule is in the zwitterion form. The molecule has an open, extended conformation mainly stabilized by an intramolecular hydrogen bond between the amino nitrogen atom and the adjacent imidazole nitrogen atom [19]. An intermolecular hydrogen bond between the $NH_3^+$ hydrogens and the nitrogen of the imidazole ring has also been found [19]. The inset of Fig. 1 shows the L-histidine molecular structure. In the unit cell of LHIS, all atoms occupy sites with $C_1$ symmetry. The 240 vibrations can be decomposed into the irreducible representations of the factor group $D_2$ as $\Gamma = 60 (A + B_1 + B_2 + B_3)$. Among these modes, three ($B_1 + B_2 + B_3$) belong to the acoustic branches. All phonons are Raman active but only phonons with $B_1$, $B_2$ and $B_3$ symmetries are infrared active.

In Fig. 2(a) we show the Raman spectra between 50 and 200 $cm^{-1}$ for several temperatures, obtained upon cooling. Here we find the first indication of a conformational change. This region encompasses the modes of the A-H···B structure [20], $\delta$(A-H···B), and low-wavenumber internal modes. The 50-160 $cm^{-1}$ region displays the lattices modes of the crystal. We make the assignments on the basis of published data for certain group wavenumbers and vibrations in amino-acid crystals. For example, the band at 170 $cm^{-1}$ is tentatively assigned as a torsional vibration $\tau(CO_2^-)$ of the $CO_2^-$ unit, an assignment based on the results for L-leucine [21] and L-alanine [22] crystals. At room temperature (285 K), in Fig. 2(a), we can clearly see five bands, which remain in the Raman spectra down to 30

K. Upon cooling the intensities of these bands are not significantly changed, although the linewidths decrease.

Lorentzian fits to their lineshapes yield the peak wavenumbers, which are plotted as functions of temperature, between 30 and 285K, in Figs. 2(b) and 2(c). The plots n the low-wavenumber region [Fig. 2(b)] are linear. The peak wavenumbers rise as the temperature drops, simply because the force constants grow as the intermolecular distance is reduced [23]. By contrast, the torsional band τ($CO_{2-}$) of the $CO_{2-}$ unit displays the more structured behavior in Fig. 2(c): (i) predominantly linear for temperatures between 165 and 285 K, and (ii) exponential for temperatures below than 161 K. The latter dependence is quite well fitted by the exponential function deduced by Balkanski et. al. [24] for optical phonons

$$\omega(T) - \omega_0 + \alpha \left(1 + \frac{2}{e^x - 1}\right) + \beta \left[1 + \frac{3}{e^y - 1} + \frac{3}{(e^y - 1)^2}\right]. \quad (1)$$

with x = $h\omega_0/2k_BT$, and y = $h\omega_0/3k_BT$. The coefficients α and β monitor the significance of the anharmonic terms. The three parameters, α, β, and $\omega_0$ on the right-hand side of Eq. (1) for τ($CO_2^-$) and other bands of L-histidine are shown in Table 1. Equation 1 indicates that, at low temperature, the scattering by the τ($CO_2^-$) mode receives contributions from three- and four-phonon processes. This contrasts with most of the other modes, which can be fitted by a straight line. The anharmonic terms arise because the oxygen atoms of the carboxylic groups directly participate in the hydrogen bonds linking the histidine molecules in the unit cell. Hydrogen bonds introduce make the potential of the carboxylic group strongly anharmonic; in other words, a realistic expression for the potential energy guiding the motion of the $CO_2^-$ unit would have to include terms of third (anharmonicity) and fourth order (stabilization of the potential). The absence of such terms partially explains the less satisfactory agreement between first-principle calculations and experimental results in amino-acid crystals and other structures chiefly stabilized by hydrogen bonds, in comparison with ionic crystals. This difficulty has been the subject of recent work [25, 26]. Another symptom of anharmonicity is the structural sensitivity to small changes in the dimension and bond angles of hydrogen bonds, which can yield distinct polymorphs of the same substance [27].

Figure 3(a) shows the Raman spectra of a L-histidine crystal in the region between 200 and 600 cm$^{-1}$. Here one expects to find the bands associated with the low-wavenumber internal vibrations of the LHIS crystal, such as the skeletal deformation modes of L-histidine, out-of-plane vibrations, and torsion of CH. The band at 244 cm$^{-1}$ is assigned to an out-of-plane vibration of CH, $\gamma$(CH) [21], while the bands at 314 and 430 cm$^{-1}$ are associated with vibrations of the skeletal structure of the L-histidine molecule, $\delta$(skel) [23, 28]. The band at 544 cm$^{-1}$ is assigned to the rocking r($CO_2^-$) of $CO_2^-$. Although it has relatively low intensity, the latter has a regular behavior, in the sense that its wavenumber is relatively stable in the investigated temperature range.

Figure 3(b) shows the Raman spectra of a L-histidine crystal in the 600-1200 cm$^{-1}$ range at different temperatures. The band at 658 cm$^{-1}$ in the room-temperature (285 K) spectrum probably corresponds to a deformation vibration of the imidazole ring and an out-of-plane vibration of NH, as suggested by Rajkumar et. al. [28]. Following previous work [21–23, 28], we have attributed the bands at 684 and 782 cm$^{-1}$ to scissoring and deformation vibrations of the carboxylate group ($CO_2^-$), respectively.

The band at 806 cm$^{-1}$ is associated with the $\delta$($CO_2^-$) bending of $CO_2^-$, as in L-cysteine [23]. The band at 854 cm$^{-1}$ is due to the wagging vibration of H-atoms at the imidazole (wag) ring [17]. The bands at 921 and 970 cm$^{-1}$ are tentatively assigned to the stretching vibrations $\nu$(CC) of CC structures [21]. In a study of L-histidine hydrochloride monohydrate Faria et al. have associated the vibration at 1064 cm$^{-1}$ with an in-plane CH deformation of the imidazole ring [17]. The band at 1091 cm$^{-1}$ is associated with a symmetric stretching $\nu_s$(CN) of CN, and the two bands at 1113 and 1179 cm$^{-1}$ can be assigned to the rocking r($NH_3^+$) of the $NH_3^+$ units [21].

Plotted as functions of temperature, the peak wavenumbers of almost all bands in the 200 – 1200 cm$^{-1}$ range are well fitted by straight lines. The two exceptions are the bands at 1113 and 1179 cm$^{-1}$, due to vibrations of the functional group $NH_3^+$, directly involved in the hydrogen bonding with a nitrogen atom of the imidazole group [19]. Figs 3(c) and 3(d) plot the peak wavenumber as functions of temperature for the lower and higher wavenumbers, respectively. Two regimes can be distinguished: (i) an approximately linear dependence between 165 and 300 K and (ii) an exponential dependence for temperatures below 161 K.

Figure 4(a) shows the temperature evolution of the Raman spectra of LHIS in the 1200-1650 cm$^{-1}$ region. One now expects to observe the bands associated with bending vibrations of the CH and $CH_2$ structures and stretching vibrations of the $CO_2^-$ units, among others. The band at 1225 cm$^{-1}$ is assigned to the twisting mode of $CH_2$ [28], and the band at 1341 cm$^{-1}$, to the deformation δ(CH) of the CH [21,22]. The bands at 1411 and 1506 cm$^{-1}$ are assigned to symmetric and asymmetric stretchings of $CO_2^-$, $ν_s(CO_2^-)$ and $ν_{as}(CO_2^-)$, respectively [5, 28].

Figure 4(b) plots the experimental wavenumber vs. temperature for modes between 1200 and 1600 cm$^{-1}$. Here we detect another evidence of the conformational change undergone by the L-histidine crystal. Approximately linear behaviors are observed for all bands. Wavenumber discontinuities are nonetheless observed for the bands at 1411 and 1506 cm$^{-1}$, associated to $ν_s(CO_2^-)$ and $ν_{as}(CO_2^-)$, respectively, as shown in Figs. 4(c) and 4(d), as well as for the band at 1575 cm$^{-1}$, associated with $ν_{as}(CO_2^-)$, as shown in Fig. 4(c). In addition to the discontinuity, the $ν_s(CO_2^-)$ wavenumber behaves anharmonically for T < 165 K. The discontinuity temperature coincides with the results for the bands associated with $τ(CO_2^-)$ and $r(NH_3^+)$. All bands showing discontinuities near 165 K are therefore related to units ($CO_2^-$ and $NH_3^+$) involved in hydrogen bonds.

Figure 5(a) shows the Raman spectra of a LHIS crystal at various temperatures in the 2800 – 3200 cm$^{-1}$ interval. The bands in this spectral region are mainly associated with stretching vibrations, of the CH, $CH_2$ and $NH_3^+$ groups, among others. The band at 2900 cm$^{-1}$ is associated with a symmetric stretching $ν_s(CH_2)$ of $CH_2$ [21, 22]. The two bands at 2968 and 2977 cm$^{-1}$ are assigned to the symmetric stretching $ν_s(CH)$ of CH or $ν_s(CH_2)$ of $CH_2$ [22, 23, 28]. The band at 3134 cm$^{-1}$ is related to a CH stretching of the imidazole ring [17]. This set of bands shows no significant temperature dependence of the wavenumbers; only the intensity of the bands originally at 2900 and 2915 cm$^{-1}$ is somewhat changed. In conclusion, we infer that only the wavenumbers of the modes related to hydrogen bonds, both intermolecular and intramolecular ones, suffer discontinuities at ~165 K. Since the bands associated with the external modes undergo no change, the transition cannot be structural. The data are only consistent with a conformational transition, possibly with movement of the imidazole group. One such movement would modify both (i) the vibrations related to $CO_2^-$ group, which has a hydrogen bond with one nitrogen of the ring, and (ii) the vibrations related to $NH_3^+$ group, which has an intramolecular hydrogen bond with a second nitrogen atom of imidazole, which explains the changes in the Raman spectra in Figs. 2-4.

The effect is similar to the phase transition in L-cysteine, revealed at low temperature by conformational changes of the zwitterion and the intermolecular contacts of the thiol group [16]. We suggest that the imidazole ring plays an important role in the phase transition in L-histidine, analogous to the role of the thiol group at the L-cysteine phase transition.

To confirm that LHIS crystals undergo a phase transition, we carried out calorimetric measurements. Figure 6 shows the differential scanning calorimetry (DSC) thermograms in the 140-300 K temperature range. The thermogram shows an enthalpic anomaly around 166 K upon heating. The peak is a consequence of the conformational transition of the unit-cell molecules. The DSC data is consistent with our interpretation of the anomalous Raman spectra and throws additional light on our observation. The thermogram shows that the hydrogen-bond network is modified at the transition temperature, as we have inferred from the changes in the bands associated with the motifs participating in the hydrogen bonds. That modification, which gives rise to a conformational change of the L-histidine molecules in the unit cell, can be understood as follows: upon cooling, the unit cell dimension diminishes and the molecules are brought together. This in turn affects the dimension of the intermolecular bond between the nitrogen of the imidazole ring in one molecule and the hydrogen of one amino group in an adjacent histidine molecule. As a consequence, the conformation of the molecules will be slightly modified.

We find it instructive to compare our low-temperature L-histidine-crystal study with the results derived from another crystal, L-histidine hydrochloride monohydrate (LHICLM), $C_6H_9N_3O_2.HCl.H_2O$ [17]. The L-histidine hydrochloride monohydrate crystal undergoes two structural phase transitions: (i) between 110 and 140K and (ii) between 60 and 80 K. The phase transitions observed in LHICLM were related to the functional group $CO_2^-$, which plays a fundamental role in the hydrogen bonds between molecules in the unit cell of the crystal, as in L-histidine (LHIS). The carboxyl goup of the amino acid, whose oxygen atoms participate in hydrogen bonds linking different molecules, might be directly related to the structural change [17]. The hydrogen bonds, including the bonds originated from the HCl and $H_2O$ molecules, are essential to link adjacent molecules. By contrast, the LHIS crystal has no HCl or $H_2O$ molecules and undergoes only one conformational phase transitions at 165 K. It is therefore possible that the greater stability of the LHIS structure, in comparison with the LHICLM structure, be a consequence of both the different distributions and the different numbers of hydrogen bonds in the unit cells of the two crystals. Another potentially important distinction is the intramolecular bond N-H...N between the $NH_3^+$ group and the imidazole ring, found in L-histidine, but not in LHICLM. Future temperature-dependent x-ray diffraction work is expected to improve our understanding of this set of phase transitions. In addition to experimental research, computational techniques, such as Density-Functional Theory calculations [29] may also help elucidate the problem, since they may provide a more precise description of the normal modes for the material investigated in this paper.

Conclusions:

We have described the temperature dependence of the vibrational properties of L-histidine crystals, down to 30 K. The structure remains stable between 165 and 300 K. However, at 165 K wavenumber discontinuities are observed in bands associated with the $NH_3^+$ and $CO_2^-$ motifs. In an attempt to identify the origin of the changes in the Raman spectra, we have carried out DSC measurements and found an enthalpic anomaly at 166 K. The anomaly in the DSC and the modifications of the Raman bands associated with the hydrogen bonds have been interpreted as due to a conformational phase transition in the L-histidine crystal at low temperatures, which modifies both the intermolecular and intramolecular H bonds.

Acknowledgments:

The authors acknowledge the financial support from the CNPq.

References

[1] E.V. Boldyreva. In Models, Mysteries, and Magic of Molecules; (Eds. J.C.A. Boeyens, J.F. Ogilvie), Springer, Berlin, 2007, p. 169.

[2] P.T.C. Freire. In: High Pressure Crystallography – From Fundamental Phenomena to Technological Applications (Org. E. Boldyreva; P. Dera), Springer, Dordrecht, 2010, p. 559.

[3] E.V. Boldyreva. J. Mol. Struct. 700, 151 (2004).

[4] K. Hasegawa, T. Ono and T. Noguchi. J. Phys. Chem. B 104, 4253 (2000).

[5] E. Deplazes, W. van Bronswijk, F. Zhu, L.D. Barron, S. Ma., L.A. Nafie and K.J. Jalkanen. Theor. Chem. Acc. 119, 155 (2008).

[6] J.G. Mesu, T. Visser, F. Soulimani and B.M. Weckhuysen. Vib. Spectrosc. 39, 114 (2005).

[7] P.A. Frey, S.A. Whitt and J. Tobin. Science 264, 1927 (1994).

[8] D. Wang, X. Zhao, M. Vargek and T.G. Spiro. J. Am. Chem. Soc. 122, 2193 (2000).

[9] D. Klug, J. Rabani and I. Fridovich. J. Biol. Chem. 247, 4839 (1972).

[10] T. Miura, T. Satoh, A. Hori-i and H. Takeuchi. Biochemistry-US 38, 11560 (1999).

[11] E.E. Weinert, C.M. Phillips-Piro, R. Tran, R.A. Mathies and M.A. Marletta. Biochemistry-US 50, 6832 (2011).

[12] V.G. Shtyrlin, Y.I. Zyavkina, E.M. Gilyazetdinov, M.S. Bukharov, A.A. Krutikov, R.R. Garipov, A.S. Mukhtarov and A.V. Zakharov. Dalton Trans. 41, 1216 (2012).

[13] D. Ringe and G.A. Petsko. Biophys. Chem. 105, 667 (2003).

[14] I.E. Paukov, Y.A. Kovalevskaya, V.A. Drebushchak, T.N. Drebushchak and E.V. Boldyreva. J. Phys. Chem. B 111, 9186 (2007).

[15] E.V. Boldyreva. Cryst. Growth Des. 7, 1662 (2007).

[16] B.A. Kolesov, V.S. Minkov, E.V. Boldyreva and T.N. Drebushchak. J. Phys. Chem. B 112, 12827 (2008).

[17] J.L.B. Faria, F.M. Almeida, O. Pilla, F. Rossi, J.M. Sasaki, F.E.A. Melo, J. Mendes Filho and P.T.C. Freire. J. Raman Spectrosc. 35, 242 (2004).

[18] J.J. Madden, E.L. McGandy and N.C. Seeman. Acta Crystallogr. B – Stru. 28, 2382 (1972).

[19] J.J. Madden, E.L. McGandy and N.C. Seeman. Acta Crystallogr. B – Stru. 28, 2377 (1972).

[20] S. Bratos. J. Chem. Phys. 63, 3499 (1975).

[21] P.F. Façanha Filho, P.T.C. Freire, K.C.V. Lima, J. Mendes Filho, F.E.A. Melo and P.S. Pizani. Braz. J. Phys., 38, 131 (2008).

[22] H. Susi and D.M. Byler. J. Mol. Struct. 63, 1 (1980).

[23] A. Pawlukojć, J. Leciejewicz, A.J. Ramirez-Cuesta and J. Nowicka-Scheibe. Spectrochim. Acta A 61, 2474 (2005).

[24] M. Balkanski, R. Wallis and E. Haro. Phys. Rev. B 28, 1928 (1983).

[25] J.D. Dunitz, A. Gavezzotti, J. Phys. Chem. B 116, 6740 (2012).

[26] C. Quarti, A. Milani, B. Civalleri, R. Orlando, C. Castiglioni, J. Phys. Chem. B 116, 8299 (2012).

[27] T.C. Lewis, D.A. Tocher, S.L. Price, Cryst. Grow. Des. 5, 983 (2005).

[28] B.J.M. Rajkumar, V. Ramakrishnan and S.A. Bahadur. J. Raman Spectrosc. 30, 589 (1999).

[29] A.M.R. Teixeira, H.S. Santos, M.R.J.R. Albuquerque, P.N. Bandeira, A.S. Rodrigues, C.B. Silva, G.O.M. Gusmão, P.T.C. Freire and R.R.F. Bento, Braz. J. Phys. 42, 180 (2012).

Table:

Table 1: Wavenumber (in cm$^{-1}$) of the peaks appearing in the Raman spectra of L-histidine (Figs. 2-5) and parameters from fitting of the experimental points. Related to the negative sign of the β coefficients and the changes in the signs of α coefficients we do not have a clear explanation.

| ω [T = 285 K] (cm$^{-1}$) | Phase I $\omega(T) = \omega_0 + \alpha T$ | | Phase II $\omega(T) = \omega_0 + \alpha\left[1+\frac{2}{e^x-1}\right] + \beta\left[1+\frac{3}{e^y-1}+\frac{3}{(e^y-1)^2}\right]$ | | |
|---|---|---|---|---|---|
| | ω$_0$ (cm$^{-1}$) | α (cm$^{-1}$/K) | ω$_0$ (cm$^{-1}$) | α (cm$^{-1}$) | β (cm$^{-1}$) |
| 78 | 89.8 | -0.042 | | | |
| 107 | 110.3 | -0.007 | | | |
| 112 | 130.8 | -0.060 | | | |
| 136 | 146.8 | -0.039 | | | |
| 170 | 182.6 | -0.044 | 175.1 | 0.311 | -0.015 |
| 244 | 244.0 | 0.001 | | | |
| 314 | 448.3 | -0.060 | | | |
| 430 | 474.3 | -0.018 | | | |
| 492 | 488.3 | 0.014 | | | |
| 544 | 544.0 | 0 | | | |
| 658 | 656.8 | 0.007 | | | |
| 684 | 685.3 | 0.002 | | | |
| 713 | 712.0 | 0 | | | |
| 782 | 791.8 | -0.031 | | | |
| 806 | 808.7 | -0.004 | | | |
| 854 | 858.9 | -0.012 | | | |
| 921 | 926.4 | -0.012 | | | |
| 970 | 964.5 | 0.017 | | | |
| 1064 | 1070.1 | -0.015 | | | |
| 1091 | 1091.4 | 0.004 | | | |
| 1113 | 1123.2 | -0.035 | 1112.1 | 5.654 | -0.906 |
| 1179 | 1190.2 | -0.038 | 1172.4 | 12.190 | -2.164 |
| 1225 | 1226.1 | -0.001 | | | |
| 1253 | 1256.3 | -0.008 | | | |
| 1275 | 1273.9 | 0.008 | | | |
| 1321 | 1322.0 | 0 | | | |
| 1341 | 1340.5 | 0.004 | | | |
| 1411 | 1420.2 | -0.032 | 1408.5 | 7.930 | -1.587 |
| 1434 | 1434.7 | 0.001 | | | |
| 1506 | 1509.5 | -0.028 | 1507.0 | -0.364 | -0.359 |
| 1550 | 1549.5 | -0.002 | | | |
| 1577 | 1580.5 | -0.025 | 1570.1 | 7.355 | -1.494 |
| 2841 | 2843.7 | -0.010 | | | |
| 2900 | 2904.6 | -0.016 | | | |
| 2915 | 2916.6 | -0.005 | | | |
| 2968 | 2965.7 | 0.015 | | | |
| 2977 | 2974.3 | 0.012 | | | |
| 3134 | 3132.5 | 0.010 | | | |
| 3150 | 3151.6 | 0.002 | | | |

(To appear in Volume 43 of the Brazilian Journal of Physics)

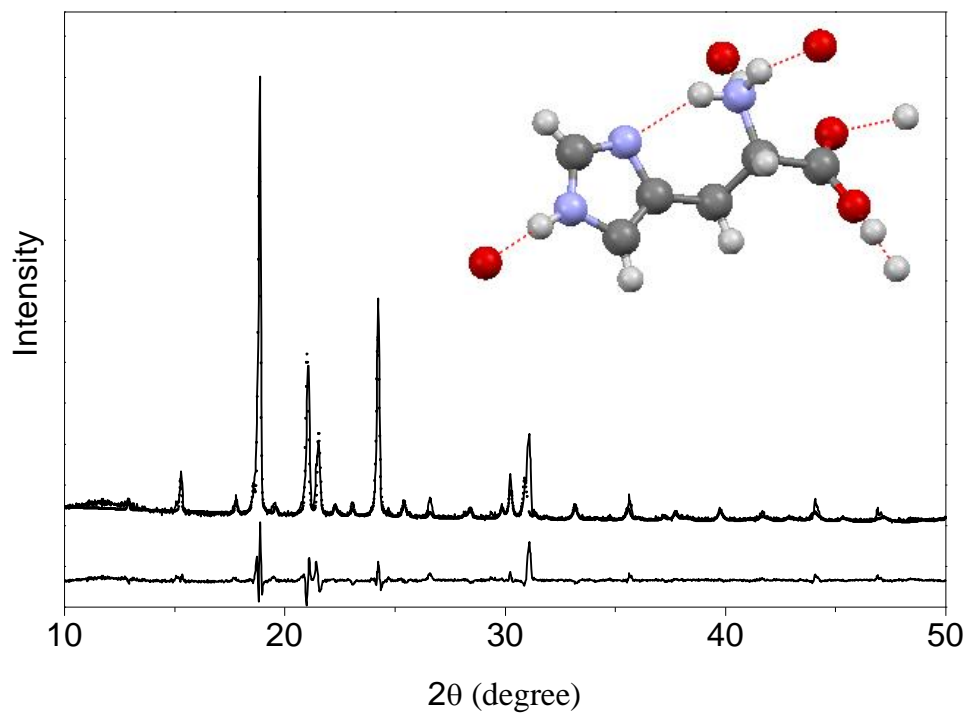

Figure 1: X-ray powder diffraction pattern for the L-histidine crystal. The solid line and crosses represent theoretical and observed intensities, respectively. The lower trace shows the difference between the calculated and experimental patterns. The inset shows the molecular structure of L-histidine.

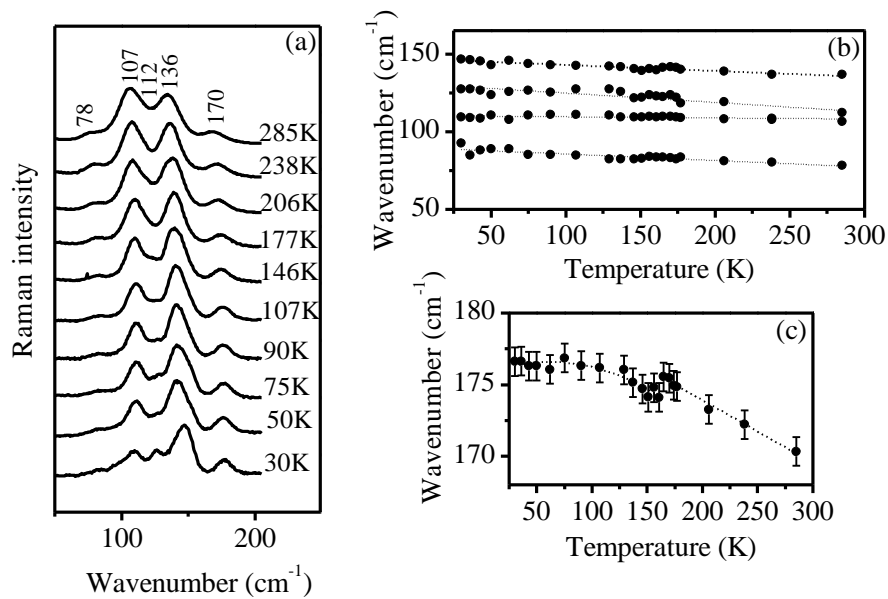

Figure 2: (a) Raman spectra of the L-histidine crystal measured at the displayed sequence of decreasing temperatures in the 50-200 cm$^{-1}$ region. (b) Wavenumbers of the bands in the 50-150 cm$^{-1}$ region as functions of temperature. (c) Temperature evolution of the wavenumber of the band associated with the torsional band of the $CO_2^-$ unit. In the T < 165K region, the solid line shows that Eq. (1) fits the experimental data well.

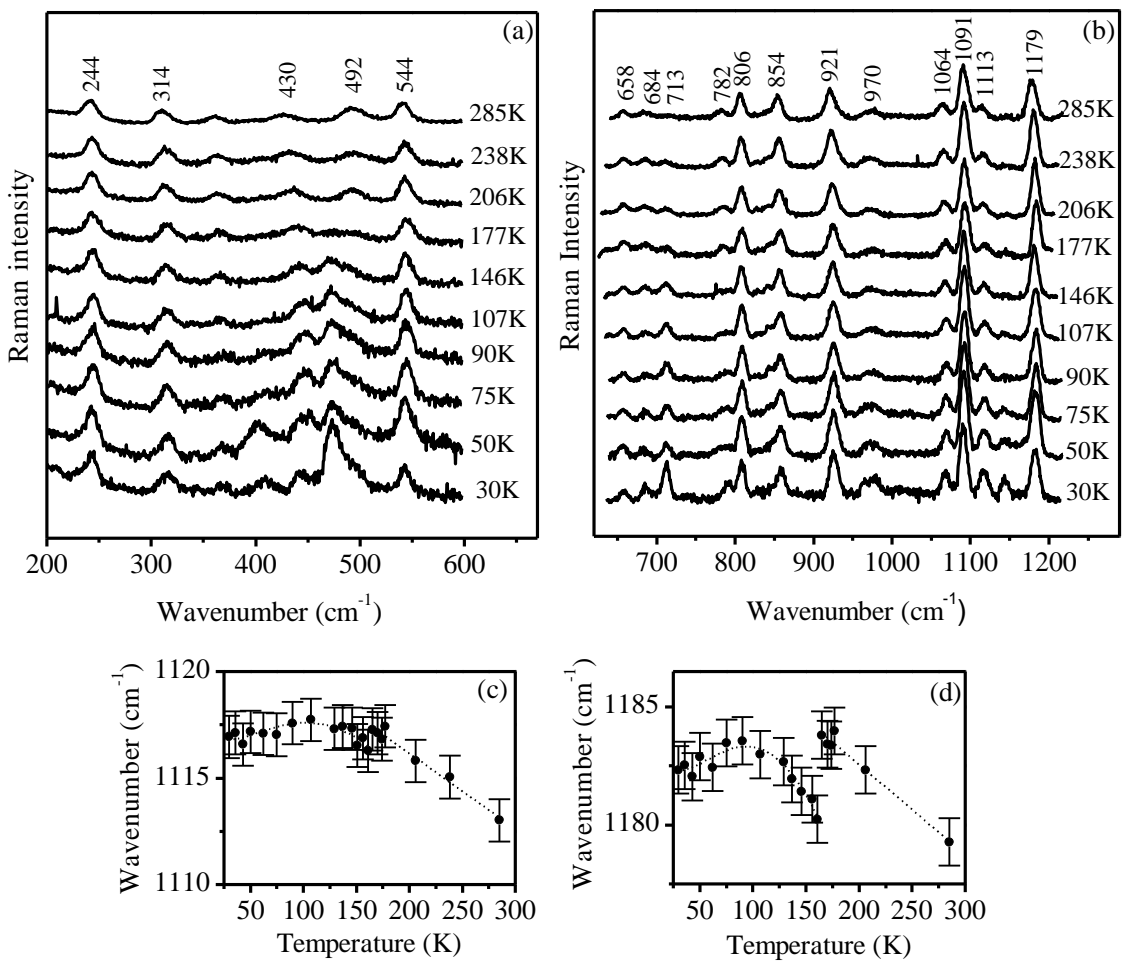

Figure 3: (a) Temperature-dependent Raman spectra of the L-histidine crystal in the 200-600 cm$^{-1}$ region, measured at the indicated sequence of decreasing temperatures. (b) Temperature-dependent Raman spectra of the L-histidine crystal in the 600-1200 cm$^{-1}$ region; (c) and (d) Thermal evolution of the wavenumber of the bands associated with vibrations of functional group NH$_3^+$. In the low-temperature range, T < 165 K, of panels (c) and (d), the experimental data are well fitted by Eq. (1), as the solid lines indicate.

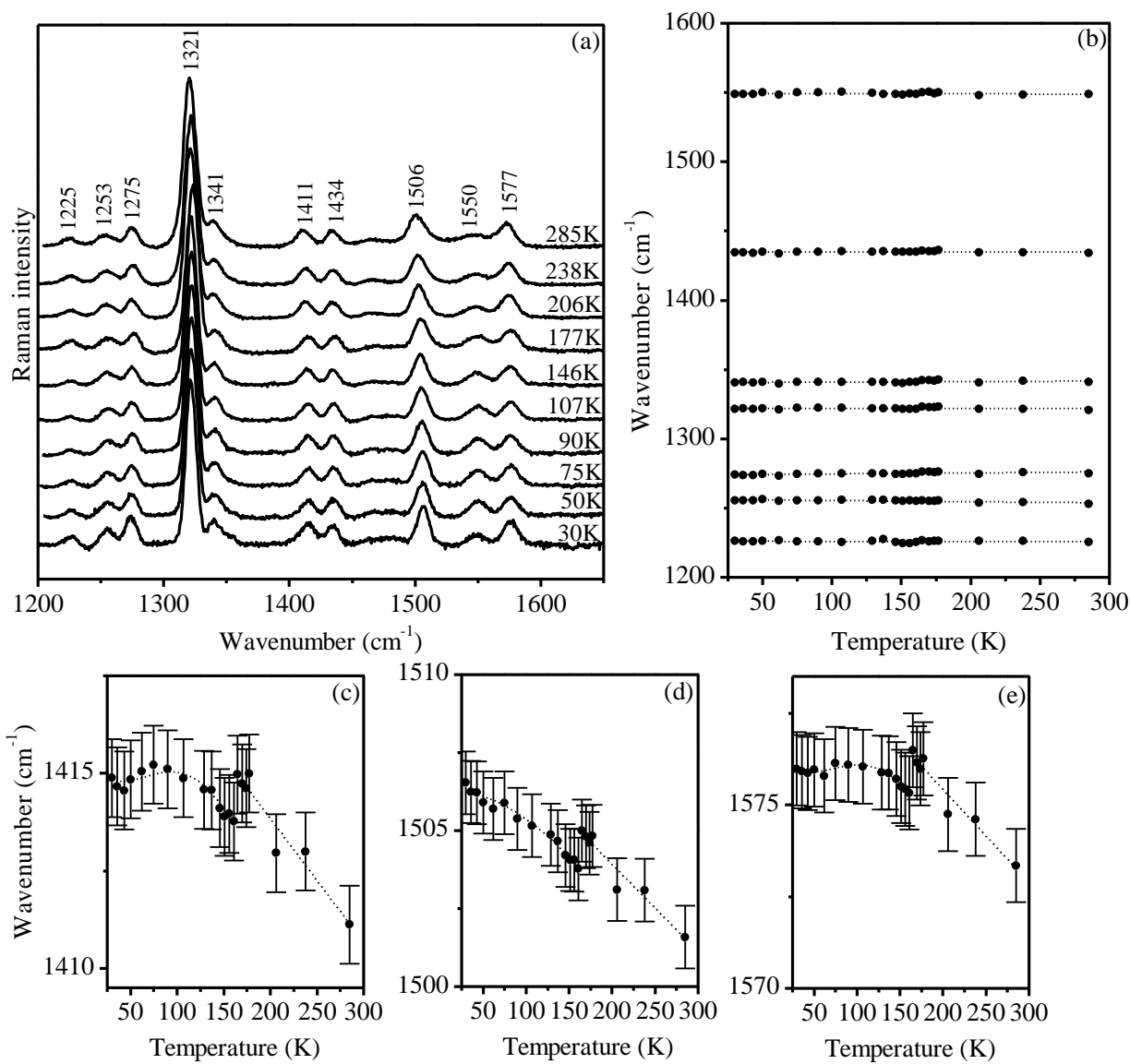

Figure 4: (a) Temperature-dependent Raman spectra of the L-histidine crystal in the 1200-1650 cm$^{-1}$ region, measured at the decreasing sequence of the indicated temperatures;.(b) Temperature dependence of the wavenumbers of the L-histidine crystal for the bands in (a). Panels (c), (d) and (e) represent the temperature evolution of the wavenumber of the bands associated with vibrations of functional group $CO_2^-$. In the low-temperature range, T <165K, the experimental data in panels (c)-(e) were fitted by Eq. (1), as indicated by the solid lines.

(To appear in Volume 43 of the Brazilian Journal of Physics)

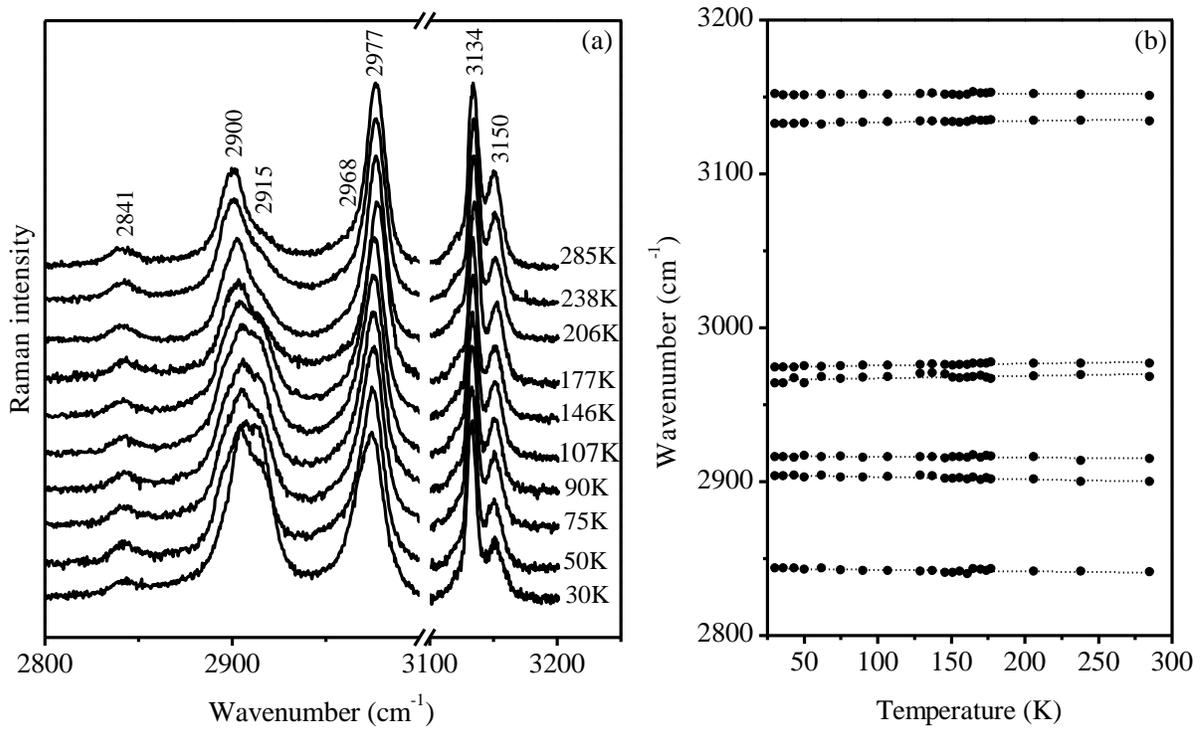

Figure 5: (a) Raman spectra of L-histidine crystal in the 2800-3200 cm$^{-1}$ spectral region for several temperatures; (b) Temperature dependence of the wavenumbers of the L-histidine crystal for the bands in (a).

(To appear in Volume 43 of the Brazilian Journal of Physics)

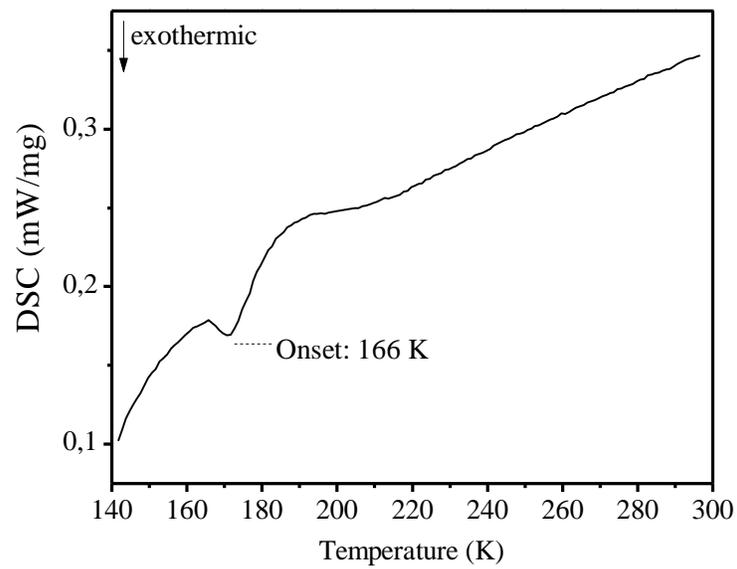

Figure 6: DSC thermogram of the L-histidine crystal in the thermal range between 140 and 300 K.